\documentclass[11pt,twoside]{article}

\usepackage{asp2006}
\usepackage{graphicx}
\usepackage{lscape}
\usepackage{multirow}
\usepackage[latin1]{inputenc}
\usepackage[OT1]{fontenc}

\bibpunct{(}{)}{;}{a}{}{,}    %% A&A fix to natbib
\def\cite#1{\citealp{#1}}     %% restore old astroncite \cite command

\makeatletter
\def\@jourvol{}
\def\cpr@year{2009}
\def\vol@title{Pathways Towards Habitable Planets}
\def\vol@author{V.\ Foresto, I.\ Ribas, D.\ Gelino (eds.)}
\makeatother
\begin{document}
%%% Fill in title
\title{Checking Stability Of Planet Orbits In Multiple-planet Systems}
%%% Fill in author names
\author{F. Malbet\altaffilmark{1}, J.\ Catanzarite, M.\ Shao, C.\
  Zhai} 
\altaffiltext{1}{now at the universit\'e de Grenoble
  J.~Fourier/CNRS, Laboratoire d'Astrophysique de Grenoble, BP 53,
  F-38041 Grenoble cedex 9, France} %%% Fill in author affiliations
\affil{Jet Propulsion Laboratory/California Institute of
  Technology, 4800 Oak Grove Drive, CA 91109 Pasadena, U.S.A.}  

\begin{abstract} %%% Abstract to run on from here.
  The SIM Lite mission will undertake several planet surveys. One of
  them, the Deep Planet Survey, is designed to detect Earth-mass
  exoplanets in the habitable zones of nearby main sequence stars. A
  double blind study has been conducted to assess the capability of
  SIM to detect such small planets in a multi-planet system where
  several giant planets might be present. One of the tools which
  helped in deciding if the detected planets were actual was an orbit
  integrator using the publicly available \texttt{HNBody} code so that the
  orbit solutions could be analyzed in terms of temporal stability
  over many orbits. In this contribution, we describe the
  implementation of this integrator and analyze the different blind
  test solutions. We discuss also the usefulness of this method given
  that some planets might be not detected but still affect the overall
  stability of the system.
\end{abstract}

%%% MAIN BODY OF TEXT GOES HERE. CONSULT "INSTRUCTIONS FOR AUTHORS USING
%%% LATEX2E MARKUP", SECTIONS 2.3-2.6 FOR HELP WITH EQUATIONS, FIGURES,
%%% AND TABLES.

%\section{}   %%% Top level section head (remove "%" symbol)
%\subsection{}   %%% Second level section head (remove "%" symbol)
%\subsubsection{}   %%% Lowest level section head (remove "%" symbol)
%\section*{}    %%% Unnumbered top level section head (remove "%" symbol)
%\subsection*{}   %%% Unnumbered second level section head (remove "%" symbol)

%\section{Introduction: context of the SIM-Lite double blind test study}

A double blind study has been conducted in the framework of the
SIM-Lite mission \citep[][see also Traub et al.\ in this
volume]{2009arXiv0904.0822T} to assess the capability of SIM to detect
such small planets in a multi-planet system where several giant
planets might be present. Five independent teams of dynamics experts
generated ensembles of hundreds of plausible planetary systems that
might form around solar-type stars. A data simulation team generated
realistic simulated astrometric and radial velocity measurement data
sets for 60 stars chosen from the SIM target list, by perturbing each
with a planetary system randomly drawn from the ensembles generated by
the dynamics teams, and then adding instrument noise. Next, four
independent analysis teams processed the simulated data sets,
detecting and fitting the orbits and masses of the planets orbiting
each of the 60 stars. Finally, the analysis teams submitted their
consensus solution for each star, chosen after cross-comparison and
discussion of the solutions, and before being allowed to see the true
solutions.

%\section{Checking the stability of orbit solutions}

We used \texttt{HNBody}, a symplectic integration package for
hierarchical N-Body systems (version 1.0.3) developed by
\citet{2002DDA....33.0802R}. It integrates the motion of particles in
self-gravitating systems where the total mass is dominated by a single
object; it is based on symplectic integration techniques in which
two-body Keplerian motion is integrated exactly. \texttt{HNBody} is
primarily designed for systems with one massive central object and has
been used previously for extrasolar planet simulations
\citep{2005ApJ...620L.111V, 2006ApJ...645.1509V}.  We used the
parameters given by the different teams, namely the mass of the
planets, their period translated in semi-major axis using the
Keplerian formula, the orbit eccentricity, the orbit inclination, the
longitude of the ascending node, the argument of periapse in years,
and for the periapse passage time, since \texttt{HNbody} only accepts
the epoch and time, we rounded the time of periapse to get the Epoch
and the time was then just the remaining decimals. In order to ensure
that we sample correctly the orbit evolution, the total integration
time has been set to 1 million years while the elementary step has
been set up to $1/20$ of the shortest period of the
system. \texttt{HNbody} computes then the relative errors in energy and
angular momentum exhibited by the integration. If the error mean and
rms are less than $10^{-4}$ over the total integration time, the
system was considered stable.

\begin{figure}[t]
  \centering
  \includegraphics[width=0.24\hsize]{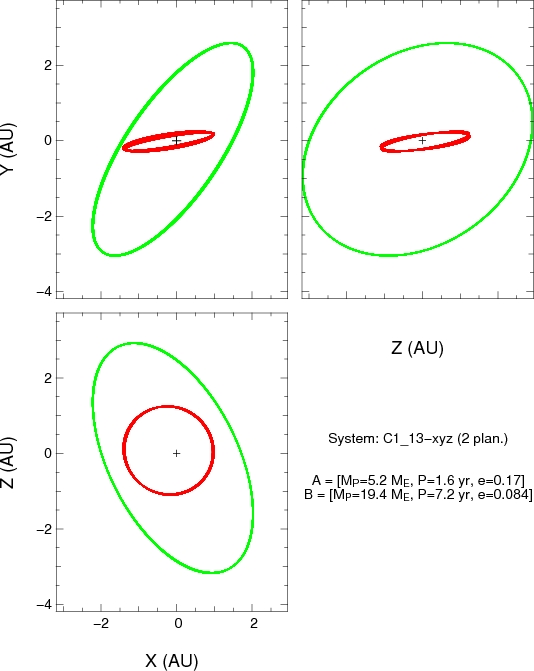}
  \includegraphics[width=0.24\hsize]{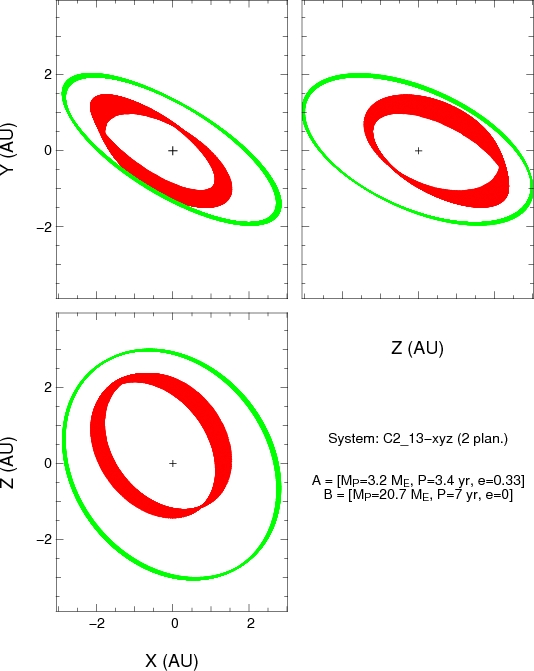}
  \includegraphics[width=0.24\hsize]{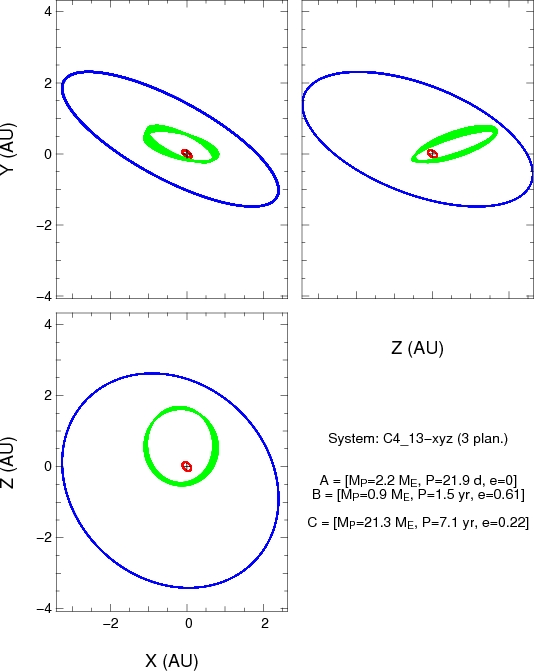}
  \includegraphics[width=0.24\hsize]{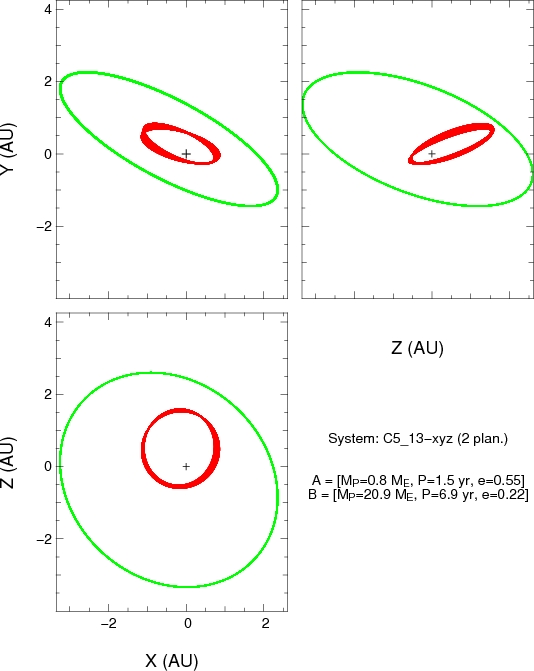}
  \caption{Planetary system \#13 of batch 2/phase 2: $xy$-, $xz$- and
    $yz$-projections of the orbits for the solutions of the 4 teams.}
  \label{fig:orbits}
\end{figure}

The plots presented in this Fig.~\ref{fig:orbits} is representative of
the temporal evolution of the orbit parameters found by the different
teams and their 3D orbits. Figure \ref{fig:orbits} display the system
for which there was intense discussion about the different solutions
and for which the stability would be decisive in choosing the
solution. This system was found stable (energy and momentum errors
less than $10^{-8}$ in $10^6$\,yrs) even though there were secular
changes eccentricity for the closest planet.

%\section{Conclusion}

The question arises whether such a stability test is appropriate for
validating a solution. Obviously, if a system misses large and massive
planets orbiting further away, then a stable system might be found
unstable because incomplete. This is the case of the system \#3 in
batch \#3 of phase 2. The unraveled solution is in fact a 9-planet
system which is stable when \texttt{HNBody} integrates the orbit over
1 million years. The reverse is true especially if only one planet is
discovered, because integration of only one planet is very likely to
be stable. Nevertheless, the orbit integration tool appeared to be
useful to choose the final solutions.

\acknowledgements %%% Text of acknowledgements runs on after this command.
FM thanks Caltech/JPL and CNRS for funding for his stay during the
year 2008-09 when this work was performed.

%\bibliographystyle{aa}
%\bibliography{pathways_poster_orbits}

\end{document}